\documentclass{aa}
\usepackage{txfonts}
\usepackage{epsfig}
\usepackage{subfigure}

\usepackage{times}
\usepackage{natbib}
\usepackage{rotating}
\bibpunct{(}{)}{;}{a}{}{,}
\begin{document}

\title{The reprocessing features in the X--ray spectrum of the NELG MCG~--5-23-16}
\author{I. Balestra, S. Bianchi, G. Matt}

 \offprints{Italo Balestra\\ \email{balestra@fis.uniroma3.it}}

\institute{Dipartimento di Fisica, Universit\`a degli Studi Roma Tre, Via della Vasca Navale 84, 00146 Roma, Italy}

\date{Received 19 August 2003/ Accepted 21 October 2003}

\authorrunning{I. Balestra, S. Bianchi, \& G. Matt}

\abstract{We present results from the spectral analysis of the Seyfert 1.9 
galaxy MCG~--5-23-16, based on $ASCA$, BeppoSAX, $Chandra$ and XMM-$Newton$ observations. 
The spectrum of this object shows a complex iron K$\alpha$ emission line, which is best modeled by a superposition of a narrow and a broad (possibly relativistic)
iron line, together with a Compton reflection component. Comparing results from all (six)
available observations, we do not find any significant variation in the flux of
both line components. The moderate flux continuum variability (about 25\% difference
between the brightest and faintest states), however, does not permit us to infer much about the location of the line-emitting material. The amount of Compton
reflection is lower than expected from the total iron line EW, implying either an iron
overabundance or that one of the two line components (most likely the narrow one) originates
in Compton-thin matter.
}

\maketitle

\section{Introduction}

X--ray spectra of type 1 Active Galactic Nuclei (AGNs) have been extensively studied by many 
satellites in the last 15 years. After $Ginga$ it has become clear that typical X--ray broad band spectra of Seyfert galaxies result from the superposition of different components. The most common features arising over the primary power law emission are two reprocessed components: the so-called Compton reflection component at high energies (above $\sim$10 keV) and the Fe K$\alpha$ emission line at 6.4 keV from neutral iron. Both are due to reflection of the primary radiation by optically thick, neutral or low ionized, circumnuclear matter. At low energies, typically below about 1 keV, a further component may arise (soft excess), the origin of which is still rather unclear.

The study of the reprocessed components and in particular of the iron line profile provides 
important information about the region from which they originate. If the line is 
produced in the innermost regions of the accretion disc, its profile must be broad, asymmetric and double-peaked due to kinematic and relativistic effects \citep[and references therein]{Fabian00}, while a narrow profile indicates an origin 
far from the nucleus, either in the torus envisaged in the Unification model \citep{Antonucci93} or in the Broad Line Regions.
 
While ASCA observations indicate that a relativistic line is a common feature in Seyfert 1s
\citep{Nandra97}, XMM-$Newton$ have so far found unambiguous evidence for relativistic
lines in an handful of objects only \citep{Wilms01, Turner02},
while finding routinely a narrow line \citep{Reeves02, Bianchi03}.

It is therefore very important for understanding the properties of the innermost regions of 
the accretion disc to search for more relativistic lines, to assess their frequency 
and strength. The fact that the relativistic line is broad, and 
therefore not easy to be separated from the continuum, and the confusing presence of the narrow component, implies that, unless the line equivalent width (EW) is very large, it should necessarily be searched for in bright sources. 

MCG~--5-23-16 is a nearby (z~=~0.0083) X--ray bright narrow emission line galaxy. Its 2--10 keV flux has varied by a factor of 4 within about 10 years: from $\sim8\times10^{-11}$ 
erg cm$^{-2}$ s$^{-2}$ in 1978 \citep{tennant83}, to $\sim2\times10^{-11}$ erg cm$^{-2}$ s$^{-2}$ in 1989 \citep{np94}, and again to a high state of $\sim9\times10^{-11}$ erg cm$^{-2}$ s$^{-2}$ in 1996 \citep{weaver98}. The nucleus is obscured by neutral matter with a column density of $\sim10^{22}$ cm$^{-2}$; no signature of a warm absorber is detected. A soft excess is present in the spectrum below about 1 keV. $RossiXTE$ detected a Compton reflection signature in this Seyfert galaxy for the first time \citep{weaver98}; $ASCA$ showed a strong (EW$\sim300$ eV) and broad Fe K$\alpha$ line with a possible complex profile \citep{weaver97}; BeppoSAX confirmed the presence of the Compton reflection component and measured an equivalent width of $\sim100$ eV for the iron line when fitted with a single gaussian \citep{per02}. More recently $Chandra$ HETGS spectra, briefly described in \citet{weaver01}, showed a narrow (FWHM $<$4\,000 km s$^{-1}$) line with EW$\sim90$ eV, while XMM-$Newton$ revealed a complex Fe K$\alpha$ line constituted by a narrow unresolved component (EW$\sim40$ eV) and a broad (FWHM $\sim40\,000$ km s$^{-1}$) component with an equivalent width of $\sim120$ eV \citep{dew03}.

In this paper we present a reanalysis of the $ASCA$, BeppoSAX, $Chandra$ and XMM-$Newton$ 
observations based on a model which includes Compton reflection and two distinct components, a narrow and a broad one, for the iron line. We will exploit the different capabilities of the various satellites (energy resolution for $Chandra$, sensitivity for XMM-$Newton$, hard X--ray coverage for BeppoSAX) to obtain informations on the location and physical properties of the emitting regions. 

\begin{table}[t]

\caption{\label{log}The log of all analysed observations, and exposure times. Values in brackets show net exposure times after filtering for particle induced flares.}

\begin{center}

\begin{tabular}{llll}
\hline
\hline
\textbf{Date} & \textbf{Mission} & \textbf{Instr.} & \textbf{T$\mathrm{_{exp}}$
(ks)} \\
\hline
11/05/1994 & $ASCA$ & \textsc{sis0-1} & 34 \\
\hline
29/11/1996 & $ASCA$ & \textsc{sis0-1} & 27 \\
\hline
 & & \textsc{lecs} & 36 \\
24/04/1998 & BeppoSAX & \textsc{mecs} & 77 \\
 & & \textsc{pds} & 33 \\
\hline
14/11/2000 & $Chandra$ & \textsc{acis-s hetg} & 76 \\
\hline
13/05/2001 & XMM-$Newton$ & \textsc{epic-pn} & 38 (6.0)\\
\hline
01/12/2001 & XMM-$Newton$ & \textsc{epic-pn} & 25 (19.1)\\
\hline
\end{tabular}

\end{center}

\end{table}

\section{\label{data}Observations and data reduction}

\subsection{ASCA}

The source was observed by $ASCA$ twice, on May 1994 and on November 1996 (Table \ref{log}). Data relative to the first observation were published by \citet{weaver97}, those relative to the second one were briefly mentioned by \citet{weaver98}. The spectra were extracted from the screened event files taken from the Goddard archive. In this paper we will only deal with the SIS0-1 spectra ($0.5-10$ keV). Data were analysed with HEAsoft 5.1 and \textsc{Xspec} 11.2.0.

\subsection{BeppoSAX}

BeppoSAX observed this source on April 1998 (Table \ref{log}). Results were published by \citet{per02}. Data reduction followed the standard procedure presented by \citet{guainazzisax99}, using HEAsoft 5.1. Spectra were extracted from regions of radius $8\arcmin$ for the LECS and $4\arcmin$ for the MECS, and analysed with \textsc{Xspec} 11.2.0. A normalization factor of 0.8 was adopted between the PDS and the MECS, appropriate for PDS spectra extracted with variable rise time threshold. The normalization of the LECS to the MECS was left as a free parameter to account for the different time coverage.

\subsection{\label{chandra}Chandra}

The $Chandra$ observation was performed on November 2000 (Table \ref{log})
with the Advanced CCD Imaging Spectrometer (ACIS-S)
and the High-Energy Transmission Grating Spectrometer (HETGS)
in place. The high flux of the source
together with a default frame time of 3.2 s resulted in a 0th order spectrum
which is strongly affected by pileup (80\% according to
WebPIMMS\footnote{http://heasarc.gsfc.nasa.gov/Tools/w3pimms.html}). Therefore,
we will only use the 1st order co-added MEG ($0.4-5$ keV) and HEG ($0.8-10$ keV) spectra in this paper. Data were reduced with the Chandra Interactive Analysis of Observations software
(CIAO 2.2.1), using the Chandra Calibration Database (CALDB 2.10). Grating
spectra were analysed with $Sherpa$ 2.2.1.

\subsection{XMM-$Newton$}

XMM-$Newton$ observed MCG~--5-23-16 twice on 13 May 2001 and 1 December 2001 for 38 ks and 25 ks respectively (Table \ref{log}). Both observations were performed in the Full Frame Mode using the blocking optical Medium Filter with the EPIC PN and MOS simultaneously operating. We will not deal with the RGS spectra in this paper because they offer poor statistics. Data were reduced with SAS 5.3.3. Events corresponding to pattern 0 were used for the PN and 0-12 for the MOS in both observations. The extracted background high energy ($\mathrm{E}>10$ keV) lightcurves clearly showed particle-induced flares, which were filtered adopting a 3 $\sigma$ threshold above the quiescent state background count rate. This resulted in net exposure times for the PN of 6.0 ks and 19.1 ks, respectively for the observations of May and December. In both observations the MOS count rates (3.5 s$^{-1}$ and 3.2 s$^{-1}$ for MOS1) are higher than the 1\% pileup limit for this instrument (0.7 s$^{-1}$, see table 3 of the XMM-$Newton$ Users' Handbook\footnote{http://xmm.vilspa.esa.es/external/xmm\_user\_support/\\documentation/uhb\_2.1/XMM\_UHB.html}), therefore we will not deal with the MOS spectra in this paper. On the other hand, the PN count rates (7.5 s$^{-1}$ and 8.0 s$^{-1}$) are just within the 1\% pileup limit (8 s$^{-1}$ for this instrument).
EPIC-PN spectra ($0.5-10$ keV) and lightcurves were extracted from a radius of 40\arcsec, and analysed with \textsc{Xspec} 11.2.0. \\

In the following, errors are at the 90\% confidence level for one interesting
parameter ($\Delta \chi^2=2.71$).

\section{Data analysis}

\subsection{Flux and spectral variability}

\begin{figure}

\centerline{\epsfig{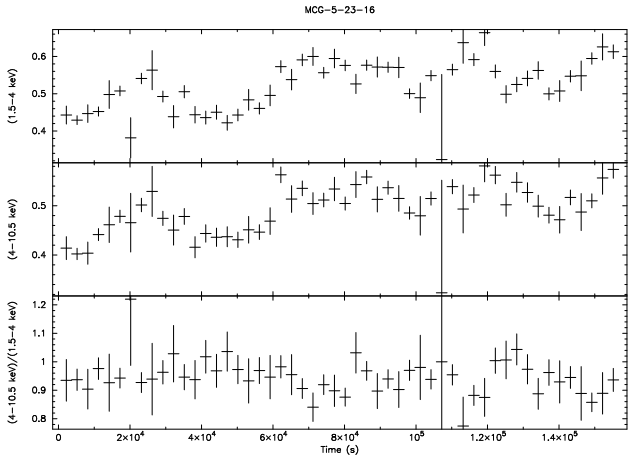}}
\caption{\label{hrmecs}BeppoSAX MECS hardness ratio ($4-10.5$ keV)/($1.5-4$ keV).}

\end{figure}

We analysed six X--ray observations of MCG~--5-23-16, covering a history
of seven years (see Table \ref{log}). The $2-10$ keV flux (averaged over an entire
observation) of the source has varied by about 25\% in about 1  year (see Table \ref{flux}). Large short-term variability is also present, as best illustrated by the BeppoSAX observation, which is the longest available (see Fig.~\ref{hrmecs}). 
Despite the remarkable flux variability, however, the hardness ratio ($4-10.5$ keV)/($1.5-4$ keV) is basically constant, implying that  no spectral variability is present at these energies. Similar temporal behaviours are also found in the other observations. 

\begin{table}[t]

\caption{\label{flux}The total flux ($2-10$ keV) as measured adopting the baseline model (described further in the text).}

\begin{center}

\begin{tabular}{cc}
\hline
\hline
\textbf{Observation} & \textbf{Flux 2-10 keV} \\
& in units of $10^{-11}$ erg cm$^{-2}$ s$^{-2}$ \\
\hline
$ASCA$ 1994 & $9.24\pm0.26$ \\
$ASCA$ 1996 & $7.45\pm0.21$ \\
BeppoSAX 1998 & $9.30\pm0.07$ \\
$Chandra$ 2000 & $9.66\pm0.11$ \\
XMM 05/2001 & $8.08\pm0.01$ \\
XMM 12/2001 & $7.12\pm0.02$ \\
\hline
\end{tabular}

\end{center}

\end{table}

\subsection{Spectral analysis}

We will only deal with data above 2.5 keV, because we are mainly interested in studying the reprocessing features, i.e. the iron lines and the Compton reflection component.
 
\begin{figure}
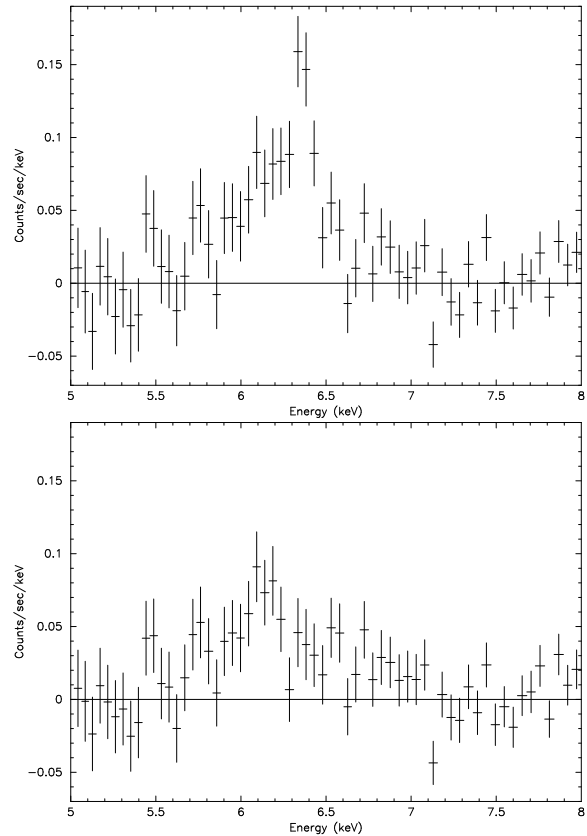


\centerline{\epsfig{figure=res_line01.ps,width=5.5cm,angle=-90}}
\centerline{\epsfig{figure=res_line02.ps,width=5.5cm,angle=-90}}
\caption{\label{iron_xmm}EPIC-PN residuals around the iron line energy for 
the second XMM-$Newton$ observation when fitting with 
a simple absorbed power law (upper panel), and with a power law plus a narrow Gaussian whose flux is fixed to $4.5\times10^{-5}$ ph cm$^{-2}$ s$^{-1}$, as explained in the text (lower panel).}

\end{figure}

Our baseline model consists of an exponentially cut-off power law and a reflection component from a neutral Compton-thick slab, isotropically illuminated by the primary radiation, subtending the solid angle $R=\Omega/2\pi$ and with an inclination angle $i$ with respect to the line of sight \citep[\textsc{pexrav} model in \textsc{Xspec}:][]{mz95}. We kept the abundances of the heavy elements $Z$ fixed to the solar value.
We also included in the model two components of the iron line to account for its complex
profile (Figure \ref{iron_xmm}), as suggested both by $ASCA$ \citep{weaver97} and
XMM-$Newton$ results \citep{dew03}: a narrow gaussian centred at 6.4 keV and a 
relativistically broadened iron line, produced by the accretion disc surrounding a Schwarzschild black hole \citep{fab89}. This choice will be further justified below.

The \textsc{diskline} model parameterizes the radial line emissivity as a power law, i.e. $\propto r^{-q}$. We fixed the line rest energy to 6.4 keV, $q=-2$ and the inner radius 
$r\mathrm{_{in}}$ to 6 $r\mathrm{_{g}}$ (last stable orbit), while we kept the outer radius $r\mathrm{_{out}}$ as a free parameter. The inclination angle of the disc $i$ has been always linked to the same value of the \textsc{pexrav} model. This is equivalent to assume that both the broad line and the reflection component are emitted in the disc. 
(For this reason we also tried to fit the reflection component with the model \textsc{refsch} instead of \textsc{pexrav}, as the former includes relativistic effects. Both for the BeppoSAX and XMM-$Newton$ spectra, however, no significant differences are found in the values of the parameters and of the $\chi^{2}$ between the two models. Therefore, and for the sake of simplicity, we will adopt the \textsc{pexrav} model throughout this work).

Finally we included a uniform cold absorber with column density N$\mathrm{_{H}}$ to take into account local absorption, in addition to the Galactic one N$\mathrm{_{g}}=8.12\times10^{20}$ cm$^{-2}$ (HEASARC W3nH\footnote{http://heasarc.gsfc.nasa.gov/cgi-bin/Tools/w3nh/w3nh.pl}).

\begin{figure}

\centerline{\epsfig{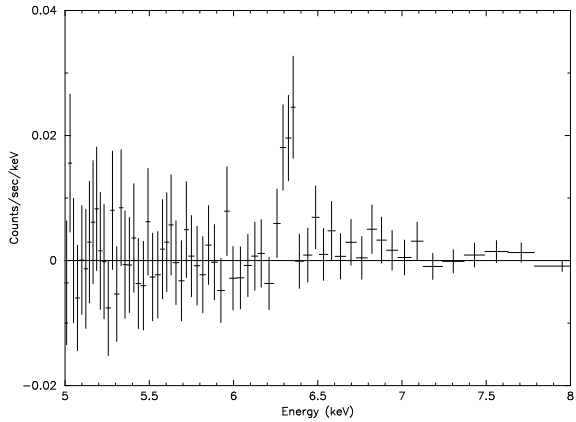}}
\caption{\label{iron_chandra}$Chandra$ HEG residuals when fitting with a
 simple absorbed power law around the iron line energy.}

\end{figure}

The analysis is made a bit complicated by the fact that, while the XMM-$Newton$ observation
is definitely the best one to constrain the iron line properties, the BeppoSAX observations
is the only capable to measure accurately enough the reflection component. We therefore
decided to adopt an iterative procedure, consisting first in estimating the inclination angle from the relativistic line in the XMM-$Newton$ second, and longer, observation, then deriving $R$, with $i$ fixed, from the BeppoSAX spectrum, and finally coming back to the XMM-$Newton$ spectrum to check if, with the so found value of $R$, the inclination angle is still the same. Once that was verified, with $R=0.45$ and $i=42^{\circ}$ obtained from such procedure, we fitted all spectra with $R$ and $i$ fixed to these values. In performing the described procedure we found $r\mathrm{_{out}}$ to be very loosely constrained. We therefore fixed it in all fits to the value giving the lowest $\chi^2$, i.e. 400~$r\mathrm{_{g}}$. 

On the other hand, if $r\mathrm{_{in}}$ is left free to vary (with $r\mathrm{_{out}}$ fixed to 400~$r\mathrm{_{g}}$), no improvement in the $\chi^2$ is found, further justifying our choice of keeping $r\mathrm{_{in}}$ fixed to the innermost stable orbit (6~$r\mathrm{_{g}}$).

The significance of the relativistic disc line is discussed in \citet{dew03}. 
Here, suffice it to say that fitting the spectrum with a narrow gaussian line results in 
$\chi^2/d.o.f.=243.4/158$; leaving the width of the line free to vary, the fit improves
giving $\chi^2/d.o.f.=198.4/157$ (significant at more than 99.99\% confidence
level according to the F-test), with $\sigma=0.28\pm0.07$ keV (corresponding to a FWHM of $30\,800\pm7\,700$ km s$^{-1}$). Such a large width is hard to explain other than being due to relativistic and kinematic effects in the innermost regions of the accretion disc \citep[see e.g.][]{Fabian00}.
Indeed, a better fit ($\chi^2/d.o.f.=192.8/157$) is obtained using a relativistic line model, with all parameters fixed to the abovementioned values but $i$, left free to vary as said above.

No significant further improvement is found adding a narrow line ($\chi^2/d.o.f.=191.5/156$). However, $Chandra$ HETGS, thanks to their excellent spectral resolution, clearly reveal the narrow iron line component at E$\mathrm{_{N}}=6.38\pm0.02$ keV (Figure \ref{iron_chandra}), which is unresolved even at the gratings resolution (FWHM$\lesssim6\,500$ km s$^{-1}$ at 99\% confidence level). Therefore, given the $Chandra$ detection of a narrow component, we decided to add it to all fits.
 
To illustrate the presence of the broad line, in Figure~\ref{iron_xmm} the EPIC-PN residuals around the iron line energy for the second XMM-$Newton$ observation are shown, when fitting with a simple absorbed power law (upper panel), and with a power law plus a narrow Gaussian 
whose flux is fixed to $4.5\times10^{-5}$ ph cm$^{-2}$ s$^{-1}$ (lower panel; see below for
the choice of this value). The broad line appears to be slightly asymmetric, 
consistently with the significant but not dramatic improvement in the $\chi^2$ when fitting with a relativistic line model rather than a broad Gaussian.

The detected 2--10 keV flux variations between the different observations 
of only  25\% at most (Table \ref{flux}) should not result 
in any measurable variability on the amount of the reflection component. This is
why we assumed $R$ to be constant and we fixed it to the BeppoSAX best fit value for all
fits. In this respect, it is reassuring that the values of $\Gamma$ so obtained are all 
consistent with one another within the error bars. The cut-off energy E$\mathrm{_{c}}$ has also been assumed to be constant (E$\mathrm{_{c}}=110$ keV). Table \ref{PX} shows the results obtained with the baseline model applied to the six 
observations considered. 

\begin{table*}

\begin{center}

\caption{\label{PX}Best fit parameters for the baseline model applied to the six observations.}

\begin{tabular}{lcccccc}
\hline
\hline
\multicolumn{1}{c}{ } & \textbf{ASCA 1994} & \textbf{ASCA 1996} & \textbf{BeppoSAX 1998} & \textbf{Chandra 2000} & \textbf{XMM 05/2001} & \textbf{XMM 12/2001} \\
\hline
N$\mathrm{_H}$ ($10^{22}$ cm$^{-2}$) & $0.89^{+0.36}_{-0.35}$ & $1.21^{+0.46}_{-0.44}$ & 
$1.57^{+0.29}_{-0.31}$ & $1.25^{+0.29}_{-0.18}$ & $1.94^{+0.38}_{-0.40}$ & $1.80\pm0.23$ \\
$\Gamma$ & $1.65^{+0.07}_{-0.09}$ & $1.69\pm0.10$ & $1.73\pm0.07$ & $1.70^{+0.06}_{-0.09}$ & 
$1.77\pm0.07$ & $1.74\pm0.04$ \\
$R$ & $0.45^*$ & $0.45^*$ & $0.45^{+0.22}_{-0.17}$ & $0.45^*$ & $0.45^*$ & $0.45^*$ \\
E$\mathrm{_{c}}$ (keV) & $110^*$ & $110^*$ & $110^{+43}_{-21}$ & $110^*$ & $110^*$ & $110^*$ \\
I$\mathrm{_{N}}$ ($10^{-5}$ph cm$^{-2}$ s$^{-1}$) & $6.8^{+1.9}_{-3.4}$ & $4.8^{+3.5}_{-3.1}$ & $7.7^{+4.3}_{-5.1}$ & $8.6\pm3.5$ & $4.1^{+2.3}_{-1.5}$ & $4.3^{+0.9}_{-1.0}$ \\
EW$\mathrm{_{N}}$ (eV) & $58^{+16}_{-29}$ & $50^{+36}_{-32}$ & $65^{+36}_{-43}$ & $70\pm28$ & $38^{+21}_{-14}$ & $45^{+9}_{-10}$ \\
I$\mathrm{_{D}}$ ($10^{-5}$ph cm$^{-2}$ s$^{-1}$) & $12.6^{+9.6}_{-7.6}$ & $9.5^{+7.9}_{-5.2}$ & $<15.4$ & $<17.6$ & $11.1^{+4.6}_{-6.2}$ & $9.9^{+2.8}_{-2.5}$ \\
EW$\mathrm{_{D}}$ (eV) & $114^{+87}_{-69}$ & $107^{+89}_{-59}$ & $<134$ & $<153$ & 
$114^{+47}_{-64}$ & $116^{+33}_{-29}$ \\
$\chi^2$/dof & 355.4/371 & 359.1/341 & 87.8/83 & 252.6/428 & 165.9/140 & 199.5/157 \\
\hline

\end{tabular}

\end{center}

NOTE.--The following parameters are fixed: iron line energy E$\mathrm{_{\alpha}}=6.4$ keV;  
disc inclination angle $i=42\degr$ (both in \textsc{pexrav} and disc line model); 
disc line emissivity $q=-2$; inner and outer disc radius 
$r\mathrm{_{in}}=6~r\mathrm{_{g}}$ and $r\mathrm{_{out}}=400~r\mathrm{_{g}}$ respectively. \\
$^*$ denotes fixed parameters.\\

\end{table*}

\begin{table*}

\begin{center}

\caption{\label{PXF}Fluxes and equivalent widths for the broad line component after the flux of the narrow line component has been fixed to the weighted mean value. Note that BeppoSAX and $Chandra$ measured fluxes are no longer only upper limits.}

\begin{tabular}{lcccccc}
\hline
\hline
\multicolumn{1}{c}{ } & \textbf{ASCA 1994} & \textbf{ASCA 1996} & \textbf{BeppoSAX 1998} & \textbf{Chandra 2000} & \textbf{XMM 05/2001} & \textbf{XMM 12/2001} \\
\hline
I$\mathrm{_{D}}$ ($10^{-5}$ph cm$^{-2}$ s$^{-1}$) & $13.5^{+6.8}_{-5.5}$ & $10.2^{+7.7}_{-5.4}$ & $10.8^{+4.4}_{-3.4}$ & $10.3^{+9.6}_{-7.7}$ & $9.8^{+5.5}_{-4.9}$ & $9.8^{+2.6}_{-2.7}$ \\
EW$\mathrm{_{D}}$ (eV) & $127^{+64}_{-52}$ & $116^{+88}_{-61}$ & $97^{+40}_{-31}$ & $91^{+85}_{-68}$ & 
$101^{+57}_{-50}$ & $114^{+30}_{-31}$ \\
$\chi^2$/dof & 356.1/372 & 359.1/342 & 88.4/84 & 255.4/429 & 165.9/141 & 199.6/158 \\
\hline

\end{tabular}

\end{center}

\end{table*}

\begin{figure}

\centerline{\epsfig{figure=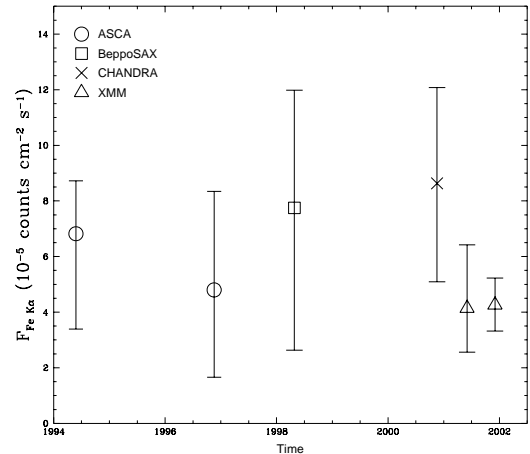,width=7cm}}
\caption{\label{flux64}Fe K$\alpha$ narrow line flux as measured adopting the baseline model.}

\end{figure}

\begin{figure}

\centerline{\epsfig{figure=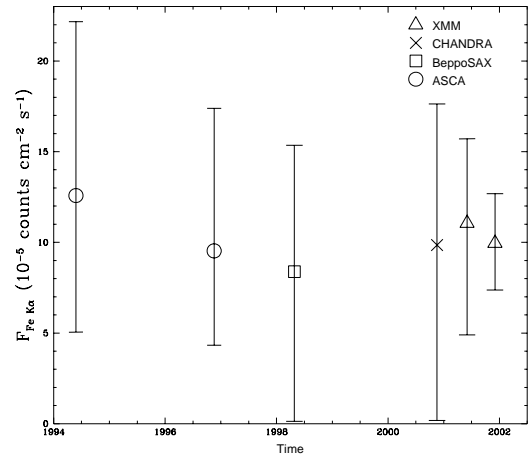,width=7cm}}
\caption{\label{broad}Fe K$\alpha$ broad line flux as measured adopting the baseline model.}

\end{figure}

\begin{figure}

\centerline{\epsfig{figure=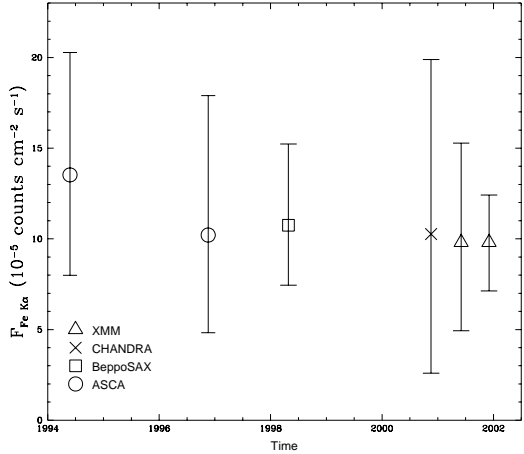,width=7cm}}
\caption{\label{broad2}Fe K$\alpha$ broad line flux as measured fixing the flux of the narrow line component to the weighted mean value ($4.5\times10^{-5}$ ph cm$^{-2}$ s$^{-1}$).}

\end{figure}

In Figures~\ref{flux64} and \ref{broad} the flux history of the narrow and broad
line components, respectively, is shown. Both components are consistent with being
constant within the, admittedly large, errors. In both cases this is not surprising, given the small flux variations in the continuum. For the narrow component, which is expected to originate in distant matter, this is also expected on theoretical grounds. To reduce the error bars on the broad component, we refitted all observations with the flux of the narrow component fixed to its weighted mean, i.e. $4.5\times10^{-5}$ ph cm$^{-2}$ s$^{-1}$. New results are summarized in Table~\ref{PXF} and illustrated in Figure~\ref{broad2}. No significant changes are found for the other parameters.

\section{\label{discussion}Discussion}

We have reanalysed the XMM-$Newton$ observations of MCG~--5-23-16, as well as previous
ASCA, BeppoSAX and $Chandra$ ones, confirming the presence of a significantly broad line already reported by \citet{dew03}. A fit with a relativistic disc profile is somewhat better than that with a broad Gaussian. Moreover, a so broad line ($\sigma=0.28$ keV) is hard to
form other than in a relativistic disc \citep[see e.g.][]{Fabian00}, which we therefore consider the most likely explanation.
A narrow line is also present, as clearly shown by $Chandra$, as well as a Compton reflection component, measured by BeppoSAX.
Exploiting the modest long term flux variability, we could confidently fit the XMM-$Newton$
spectrum adding these two further components (with the parameter $R$ describing the relative
amount of reflection component kept fixed to the value found by BeppoSAX).  

The detection of a broad, likely relativistic line in this source is important. Predicted on theoretical grounds, relativistic lines were found to be common in ASCA spectra of AGN \citep{Nandra97}.
However, \citet{Lub01} showed that at least part of the lines could be 
ascribed to a narrow component from distant matter, also to be expected 
on the basis of Unification Models \citep[e.g.][]{Ghi94}. $Chandra$ and XMM-$Newton$
have indeed confirmed the almost ubiquity of the narrow component, while relativistic
lines have been unambiguously detected so far only in a few cases 
\citep[e.g.][]{Wilms01, Turner02}. 
The discovery of a relativistic line in MCG~--5-23-16 adds therefore one more
case to a still short list. It is worth remarking that the relativistic lines
have been observed so far in sources in which either the EW of the line is very large, 
as in the case of MCG~--6-30-15 and NGC~3516 \citep{Turner02}, or that are very bright, like MCG~--5-23-16. The presence of the narrow component, in fact, could make difficult to detect the broad component in not very bright sources, if its EW is modest. Even if in some cases, e.g. NGC~5506 \citep{Matt01, Bianchi03}, NGC~7213 \citep{b03}, NGC~5548 \citep{Pounds03} and NGC~4151 \citep{Schurch03}, relativistic lines with significant equivalent widths are definitely absent in high signal-to-noise XMM-$Newton$ spectra, the actual frequency of relativistic lines is still to be assessed.

The amount of reflection component, $R=0.45$, is low if compared with the total EW of the
lines, which is about 170 eV \citep[e.g.][]{matt91, gf91}. One possible
explanation is iron overabundance \citep{Matt97}; alternatively, one of the two line components may originate in Compton-thin matter. The obvious candidate is the narrow component (the broad one being likely emitted by the accretion disc), which therefore could be associated with the BLR rather than the molecular torus. (It is worth noting that the column density of the absorber is too low to make it a good candidate for producing the line
\citep[e.g.][]{Matt03}). This hypothesis cannot be directly tested, because 
MCG~--5-23-16 is a NELG and therefore the BLR is supposed to be obscured in the optical
band, and therefore no BLR line width, to be compared with the upper limit derived from $Chandra$, is available. Alternatively, the line can be emitted in a Compton-thin torus. 
If indeed the narrow line is emitted in Compton-thin matter, no evidence for a 
Compton-thick torus is present in this source, similar to the case of NGC~7213 \citep{b03}.

\acknowledgement
We thank G.C. Perola for useful discussions on BeppoSAX data analysis, Kazushi Iwasawa for his help on reducing $ASCA$ data and the anonymous referee for helpful comments and suggestions. We acknowledge financial support from ASI.

\bibliographystyle{aa}
\bibliography{sbs}

\end{document}